\documentclass[twocolumn,showpacs,preprintnumbers,amsmath,amssymb,prb]{revtex4}

\usepackage{graphicx}
\usepackage{dcolumn}
\usepackage{bm}

\begin{document}
\draft
\title
{
Weak ferromagnetism of La${}_{1.99}$Sr${}_{0.01}$CuO${}_4$ thin films: 
evidence for removal of corrugation in CuO${}_2$ plane 
by epitaxial strain
}
\author{I. Tsukada}
\email{ichiro@criepi.denken.or.jp} 
\affiliation{
Central Research Institute of Electric Power Industry, 
2-11-1 Iwado-kita, Komaeshi, Tokyo 201-8511, JAPAN 
}
\date{\today}

\begin{abstract}
The weak ferromagnetism of La${}_{1.99}$Sr${}_{0.01}$CuO${}_4$ epitaxial 
thin films is investigated using magnetoresistance measurement. 
While a steplike negative magnetoresistance associated with the 
weak ferromagnetic transition is clearly observed in the films grown 
on YAlO${}_3$(001), it is notably suppressed in the films 
grown on SrTiO${}_3$(100) and 
(LaAlO${}_3$)${}_{0.3}$(SrAl${}_{0.5}$Ta${}_{0.5}$O${}_3$)${}_{0.7}$(100), 
and almost disappears in films grown on LaSrAlO${}_4$(001). 
The strong suppression of the steplike magnetoresistance provides 
evidence that the CuO${}_2$ planes are much less corrugated in thin films 
grown on tetragonal substrates, particularly on LaSrAlO${}_4$(001), 
than in bulk crystals. 
\end{abstract}

\maketitle

\narrowtext

Epitaxial growth of La${}_{2-x}$Sr${}_x$CuO${}_4$ (LSCO) and 
La${}_{2-x}$Ba${}_x$CuO${}_4$ (LBCO) superconductors 
has been gathering much attention since a remarkable 
increase in $T_c$ was reported for films grown on LaSrAlO${}_4$(001).
\cite{Sato1,Sato2,Locquet1} 
In these studies, in-plane compressive and out-of-plane tensile strains 
applied to the lattice were found to be favorable for the increase of $T_c$. 
As is widely known, bulk crystals of LSCO have two phases: 
high-temperature tetragonal (HTT, $I4/mmm$) and low-temperature orthorhombic 
(LTO, $Bmab$) phases. 
In the LTO phase, a staggered rotation of the CuO${}_6$ octahedra 
gives rise to the characteristic corrugation in the CuO${}_2$ plane. 
Considering that superconductivity occurs in this LTO-phase stable region 
for bulk {\it unstrained} crystals, 
we expect that the epitaxial strain affects the superconducting 
properties through a modification in the corrugated structure. 
However, it seems difficult to directly investigate a such microscopic 
structure in thin-film samples; 
we usually apply a neutron diffraction method to a bulk crystal 
for such purpose, but it may not apply to a thin-film sample 
because of its too small volume. 
Thus far, there has been no report on the true role of epitaxial strain 
of LSCO and LBCO thin films.

An indirect but prospective approach to the corrugation in the CuO${}_2$ 
plane is to study an antiferromagnetic sample instead of 
a superconducting one, because the corrugation is strongly coupled 
to antiferromagnetism and weak ferromagnetism of LSCO. 
In the LTO phase, the corrugated structure of the CuO${}_2$ plane allows 
Dzyaloshinskii-Moriya (DM) interactions to work between the nearest-neighbor 
spins on Cu${}^{2+}$ ions.
\cite{Thio1} 
The DM interactions consequently induce cooperative spin canting, 
giving rise to a weak ferromagnetic (WF) moment perpendicular to the 
CuO${}_2$ plane, 
which alternately changes its direction along the $c$ axis. 
Since the coupling of the WF moments on adjacent layers is rather weak, 
one can drive a WF transition by applying magnetic field along the $c$ axis; 
above the critical field, the WF moments point in the same direction 
at every CuO${}_2$ plane. 
Thio {\it et al.} have reported that the out-of-plane resistance shows 
a steep decrease at the critical field,
\cite{Thio1} 
indicating a strong coupling of electronic transport and magnetic ordering. 
Recently, Ando {\it et al.} have discovered that the in-plane magnetoresistance 
also exhibits a characteristic steep decrease across the WF transition.
\cite{Ando2} 
The in-plane magnetoresistance, therefore, can be a sensitive probe of 
the weak ferromagnetism that is strongly coupled with the corrugated 
structure of the CuO${}_2$ plane. 
In this paper, we report the in-plane magnetoresistance for the 
antiferromagnetic (AF) La${}_{1.99}$Sr${}_{0.01}$CuO${}_4$ thin films 
epitaxially grown on four different substrates, and discuss how the 
corrugation of the CuO${}_2$ plane is modified by the epitaxial strain.

La{}$_{1.99}$Sr${}_{0.01}$CuO${}_4$ thin films were prepared 
by pulsed-laser deposition (KrF excimer, $\lambda$ = 248~nm). 
Substrate temperature was kept at 800${}^{\circ}$C during the deposition 
in an atmosphere of 30~mTorr oxygen. 
This growth condition is the same as that used for a previous report.
\cite{Lavrov2} 
Under this growth condition we can grow superconducting LSCO thin films with 
very low residual resistivity implying few crytallographic imperfections. 
The laser repetition rate was set at the lowest value ($f$ = 1~Hz) 
to properly apply strain to the films. 
A polycrystalline sintered target of La{}$_{1.99}$Sr${}_{0.01}$CuO${}_4$ 
was prepared by a solid-state reaction method. 
At this Sr concentration, we expect that the AF long-range order 
appears at $T \approx$ 200~K in a bulk crystal.
\cite{Lavrov1} 
The actual chemical composition of the grown film became slightly La-rich 
(La:Sr:Cu = 2.28:0.01:1.00), 
but this deviation does not significantly affect the antiferromagnetism 
of the sample. 
We used four substrates: 
orthorhombic YAlO${}_3$(001) has a rectangular surface symmetry, 
while tetragonal LaSrAlO${}_4$(001), cubic 
(LaAlO${}_3$)${}_{0.3}$(SrTa${}_{0.5}$Al${}_{0.5}$O${}_3$)${}_{0.7}$(100), 
and cubic SrTiO${}_3$(100) have a square surface symmetry. 
Hereafter, these substrates are referred to as YAP 
(Yttrium Alminum Perovskite), LSAO, LSAT, and STO, respectively. 
To avoid confusion, we follow the axis notation of the LTO structure of LSCO. 
Film thickness was set at approximately $t$ = 1800{\AA}. 
After patterning the films for four-terminal measurements, 
they were carefully annealed at 600${}^{\circ}$C in helium to remove 
extra oxygen following the procedure for lightly doped bulk crystals.
\cite{Lavrov1} 
Since the extra oxygen can easily induce superconductivity, 
as discussed by Sato {\it et al.}
\cite{Sato1} 
and Bozovic {\it et al.},
\cite{Bozovic1} 
this postannealing in helium is crucial for the present experiments. 
The in-plane resistivity was measured under the magnetic field up to 10~T by 
the Physical Properties Measurement System (PPMS, Quantum Design).

\begin{figure}
\includegraphics*[width=85mm]{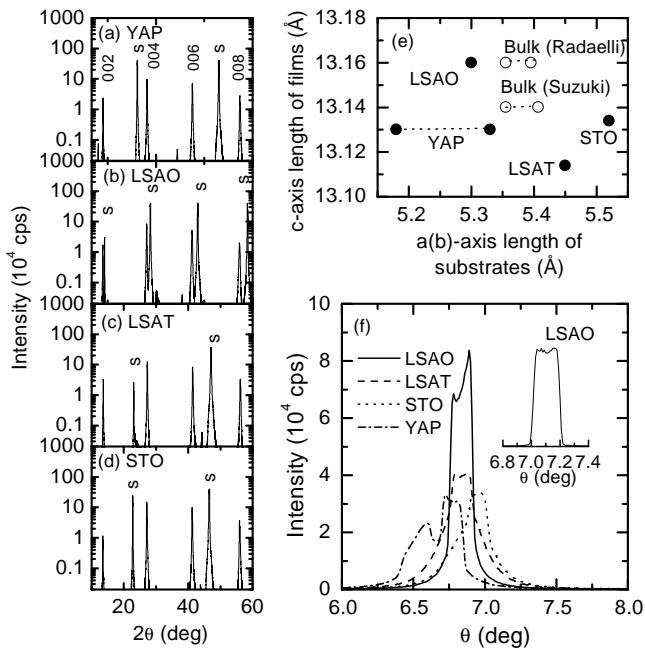}
\caption{
(a)-(d) X-ray diffraction of the films taken by $\theta$-2$\theta$ 
goniometer. Reflections from the substrates are indicated by s. 
(e) Relationship between the lattice parameters of substrates and 
the $c$-axis lengths of the films. 
The data for bulk samples of La${}_2$CuO${}_4$ are taken from Refs.~9 
and 10. 
(f) The Rocking curve of the 002 reflection. Inset shows that of 
LSAO substrate. 
}
\label{Fig.1}
\end{figure}

The x-ray diffraction shows that all the films are highly $c$-axis oriented 
as shown in Figs.~\ref{Fig.1}(a)-\ref{Fig.1}(d). 
The $c$-axis length estimated from all of the observed 00$l$ ($l$ : even) 
reflections is plotted as a function of the $a$($b$)-axis length of substrates 
in Fig.~\ref{Fig.1}(e) with the data of ceramic samples of La${}_2$CuO${}_4$.
\cite{Radaelli1,Suzuki1} 
Although the $c$-axis length does not show monotonic behavior 
at first glance, this variation is consistent with the results reported 
by Sato {\it et al.} for La${}_{1.85}$Sr${}_{0.15}$CuO${}_4$ films,
\cite{Sato1} 
{\it i.e.,} 
a significant change in the $c$-axis length is observed only 
in the films with sufficiently small lattice misfit. 
The marked expansion or compression of the $c$ axis is observed 
for the film grown on LSAO or LSAT. 
Their in-plane lattice parameters are very close to those of LSCO, 
which results a remarkable expansion or compression of the $c$-axis length. 
In contrast, the films gown on YAP and STO, where the lattice misfits 
are much larger, do not feel the strain fully, 
and consequently, the $c$-axis length approaches a certain value 
that is determined only by thermodynamics. 
We actually observed that the $c$-axis lengths of these films are almost equal.

\begin{figure}[b]
\includegraphics*[width=85mm]{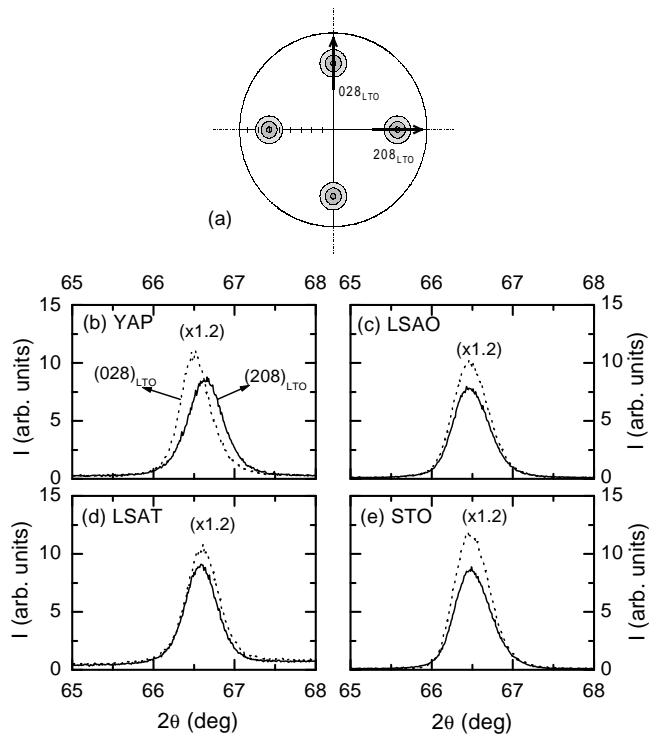}
\caption{
(a) Schematic of 208 and 028 reflections in 
the pole-figure configuration. 
The scan directions of the peaks in (b)-(e) are indicated by arrows. 
(b)-(e) 208 (solid line) and 028 (dotted line) reflections of the films 
grown on YAP(001), LSAO(001), LSAT(100), and STO(100). 
The intensity of the 028 reflection is multiplied by 1.2. 
}
\label{Fig.2}
\end{figure}

In order to see more detail, 
we compared the rocking curve around the 002 refelection 
as shown in Fig.\ref{Fig.1}(f). 
The 002 reflection is the lowest-index diffraction, and 
considered to be the most sensitve to wide-area crystallographic quality. 
We see that the peak height of the film on LSAO is 
twice as high as that of the others, 
which indicates the better orientation than the others. 
The more important feature is, however, a shape of the peak; 
The peak width is the narrowest for the film on LSAO. 
The splitting at the peak top is due to the finite mosaicness 
of the LSAO substrate (see the inset), 
and thus, the peak width of each domain is considered to be much narrower 
than the width shown in Fig.~\ref{Fig.1}(f). 
Another important feature is that this peak does not have a broad tail, 
which is only the case of the film on LSAO. 
The absence of the tail indicates that the epitaxial strain is applied 
entirely through the film. 
The contrasting behavior is observed in the films grown on LSAT and STO, 
where the broad tail is observed in their rocking curves. 
The peak structures of these films look like a superposition of 
a narrow intense and a broad less-intense peaks. 
The presence of the narrow intense peak indicates that a prat of the 
film is well strained like the film on LSAO. 
However, the accompanuing broad peaks suggests that the lattice relaxation occurs 
in another part of the films. 
In general, when lattice relaxation proceeds according to the film growth, 
the film begins to be divided into domains with slightly different 
orientation when the film can no longer hold the strain. 
Thus, the rocking curves of these films show that the films grown on LSAT 
and STO may consists of lower fully-strained layers and upper less-strained 
(relaxed) layers. 
The rocking curve of the film grown on YAP looks much more complicated. 
However, if you looks at the each peak, you can see that 
the peak width is comparable to the others. 
From this rocking curve, we expect that the lattice relaxation begins almost 
at the initial stage of the film growth, and thus, cracks also start to grow 
from the film-to-substrate interface. 
As will be shown later, we observe the highest resistivity in the film grown on YAP, 
which is consistent with this expectation.

\begin{figure}
\includegraphics*[width=80mm]{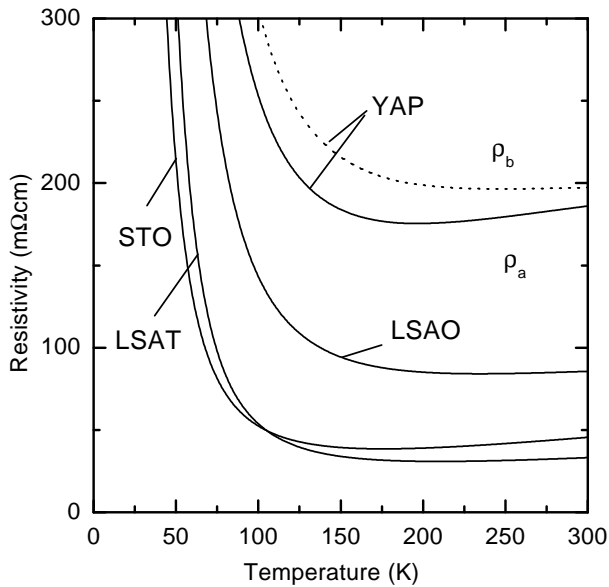}
\caption{
Temperature dependence of the in-plane resistivity of the films. 
For the film grown on YAP, the resistivity is anisotropic along two orthogonal 
directions (solid and dotted lines), while such an anisotropy is 
absent in the other films. 
}
\label{Fig.3}
\end{figure}

Next, we confirmed the in-plane crystallographic symmetry. 
For this purpose, we used a pole-figure goniometer to scan both 
the 208 and 028 reflections. 
Figures~\ref{Fig.2}(b)-\ref{Fig.2}(e) show the single scans of the 
208 and 028 reflections. 
The scan direction is indicated by an arrow in Fig.~\ref{Fig.2}(a). 
For the film grown on YAP, the 208 and 028 reflections appear at different 
angles indicating the orthorhombic symmetry, and the in-plane orientation 
is determined as YAP [100] $\parallel$ LSCO [100], 
which may allow us to expect that the corrugation survives 
in the CuO${}_2$ plane. 
On the other hand, the two reflections appear at the same angle for the 
films grown on LSAO, LSAT, and STO. 
However, these results do not immediately mean that these films are 
really tetragonal. 
For example, Bi${}_2$Sr${}_2$Ca${}_{n-1}$Cu${}_n$O${}_{2n+4}$ is orthorhombic 
and does not become tetragonal even when grown on STO(100), 
but it exhibits a twin structure composed of orthorhombic domains.
\cite{Eckstein1} 
Thus, we cannot judge whether the corrugation is actually removed 
from these three films as this stage, and therefore, 
the magnetoresistance measurement becomes important.

The in-plane resistivity of all films exhibit an insulating behavior 
upon approaching $T$ = 0~K, as shown in Fig.~\ref{Fig.3}. 
In contrast to the resistivity data reported for superconducting samples,
\cite{Sato1,Sato2,Locquet1} 
the films grown under the compressive strain (YAP and LSAO) show a higher 
resistivity than those grown under the tensile strain (LSAT and STO). 
The film grown on YAP is orthorhombic, as was shown before, 
and has resistivity anisotropy along the $a$ and $b$ axes; 
$\rho_a$ is always lower than $\rho_b$ for this film, and the temperature 
where the resistivity shows a minimum value is also lower for $\rho_a$ 
than for $\rho_b$. 
Such different temperature dependences are qualitatively consistent with those 
for bulk crystals with $x \geq$ 0.02,
\cite{Ando1} 
which supports our identification of the $a$- and $b$-axis directions. 
The other three films show almost no in-plane anisotropy.

\begin{figure}
\includegraphics*[width=85mm]{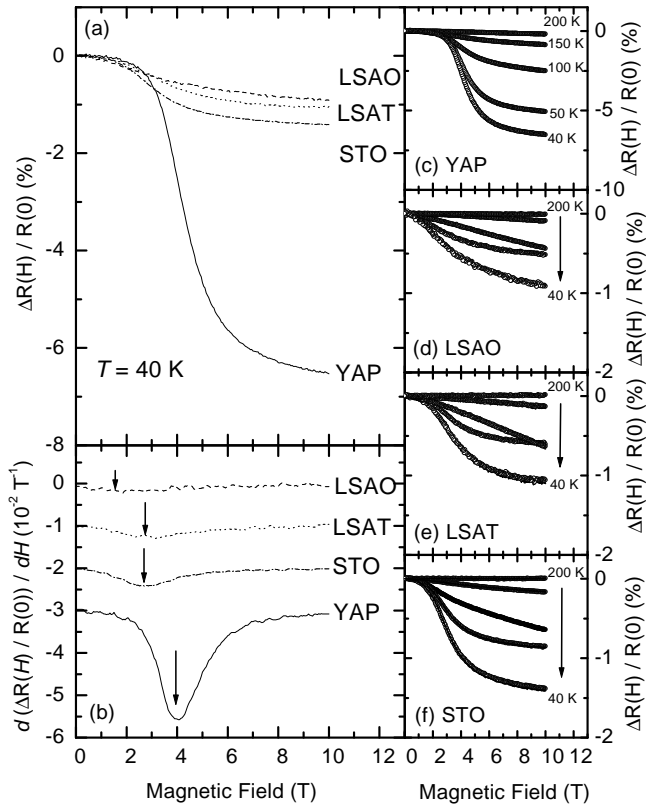}
\caption{
(a) In-plane magnetoresistance of La${}_{1.99}$Sr${}_{0.01}$CuO${}_4$ 
thin films measured at $T$ = 40~K. 
The large steplike negative magnetoresistance associated with 
the WF transition, which is significant in the film grown 
on YAP, almost disappears in the films grown on LSAO, LSAT, and STO. 
(b) The first derivative of the magnetoresistance shown in (a). 
(c)-(f) The temperature variations of the in-plane magnetoresistance 
for each film.
}
\label{Fig.4}
\end{figure}

\begin{figure}
\includegraphics*[width=80mm]{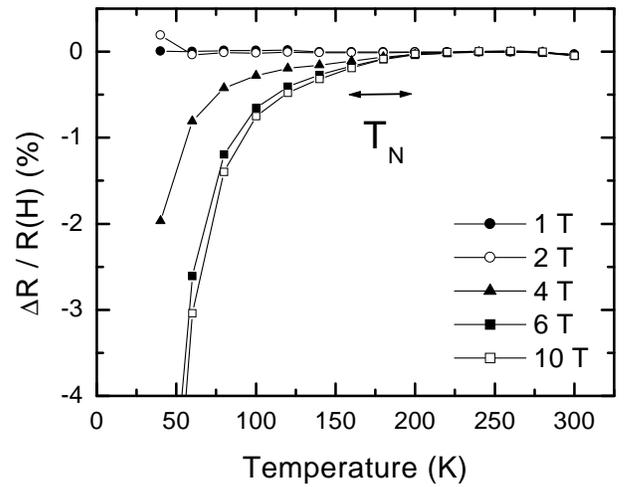}
\caption{
Temperature dependence of the magnetoresistance for 
La${}_{1.99}$Sr${}_{0.01}$CuO${}_4$ films grown on YAP (001) 
measured at $H$ = 1, 2, 4, 6, and 10~T.
}
\label{Fig.5}
\end{figure}

After characterizing the films by x-ray diffraction and 
resistivity measurements, we measured the in-plane magnetoresistance. 
Figures~\ref{Fig.4}(a) and \ref{Fig.4}(b) show the magnetoresistance 
measured at $T$ = 40~K and its first derivative, respectively, 
in which we can easily see a marked difference 
between the film grown on YAP and those grown on the others. 
The film grown on YAP exhibits a clear steplike magnetoresistance, 
like bulk crystals do.
\cite{Ando2} 
As shown in Fig.~\ref{Fig.4}(c), 
${\Delta}R(H)/R(0)$ at $T$ = 200~K is almost field-independent 
up to $H$ = 10~T, 
while below $T$ = 150~K, ${\Delta}R(H)/R(0)$ shows an apparent decrease 
above approximately $H$ = 2~T; 
a steep decrease begins at $H$ = 2~T and almost ends at $H$ = 6~T. 
For convenience, we will define the WF transition field as the field 
where the first derivative shows a minimum. 
According to this definition, $H_{WF}$ = 4~T is obtained for this film 
as indicated by an arrow in Fig.~\ref{Fig.4}(b). 
From the magnetoresistance data, we can roughly estimate the N\'eel 
temperature ($T_N$) for this film by plotting the temperature dependence of 
${\Delta}R(H)/R(0)$ at several fields (Fig.~\ref{Fig.5}). 
${\Delta}R(H)/R(0)$ is almost zero at $H < H_{WF}$ (1 and 2~T) 
through the temperature range below 300~K, 
while the deviation to negative values becomes apparent below $T$ = 200~K 
when the field is higher than $H_{WF}$ (6 and 10~T). 
As a result, the AF long-range order is considered to set in 
at approximately $T$ = 200~K. 
This assignment is consistent with $T_N$ reported for bulk crystals.
\cite{Lavrov1}

The other three films grown on tetragonal surfaces show contrasting behavior. 
The steplike magnetoresistance behavior is strongly suppressed 
in comparison with the films grown on YAP. 
At $T$ = 40~K and $H$ = 10~T, ${\Delta}R(H)/R(0)$ reaches only -1.1{\%} (LSAO), 
-0.9{\%} (LSAT), and -1.4{\%} (STO), 
which are far smaller than that observed for the film grown on YAP. 
Nevertheless, one can find traces of the WF transition in the films 
grown on LSAT and STO. 
Figure~\ref{Fig.4}(b) shows that the WF transition still occurs at 
$H_{WF} {\approx}$ 2.5~T for these two films, 
even though the critical field is lower than that for the film on YAP. 
In these films, we may expect that $T_N$ shifts to lower temperatures. 
As shown in Fig.~\ref{Fig.4}(e), the magnetoresistance of the film 
on LSAT maintains a convex upward-field dependence even at $T$ = 100~K 
and a trace of the steplike behavior appears only below $T$ = 50~K, 
which suggests that the $T_N$ of this film is found between 50 and 100~K.

The weak ferromagnetism is more strongly suppressed in the film 
grown on LSAO, and the first derivative of the magnetoresistance 
becomes almost field-independent. 
We can no longer obtain clear $H_{WF}$ from Fig.~\ref{Fig.4}(b). 
If one carefully analyzes Fig.~\ref{Fig.4}(b), 
$H_{WF}$ is still discernible around $H$ = 1.5~T. 
However, this value is even lower than the $H_{WF}$'s of the films 
grown on LSAT and STO, 
and we conclude that the weak ferromagnetism is most strongly suppressed 
in the film grown on LSAO.

\begin{figure}
\includegraphics*[width=85mm]{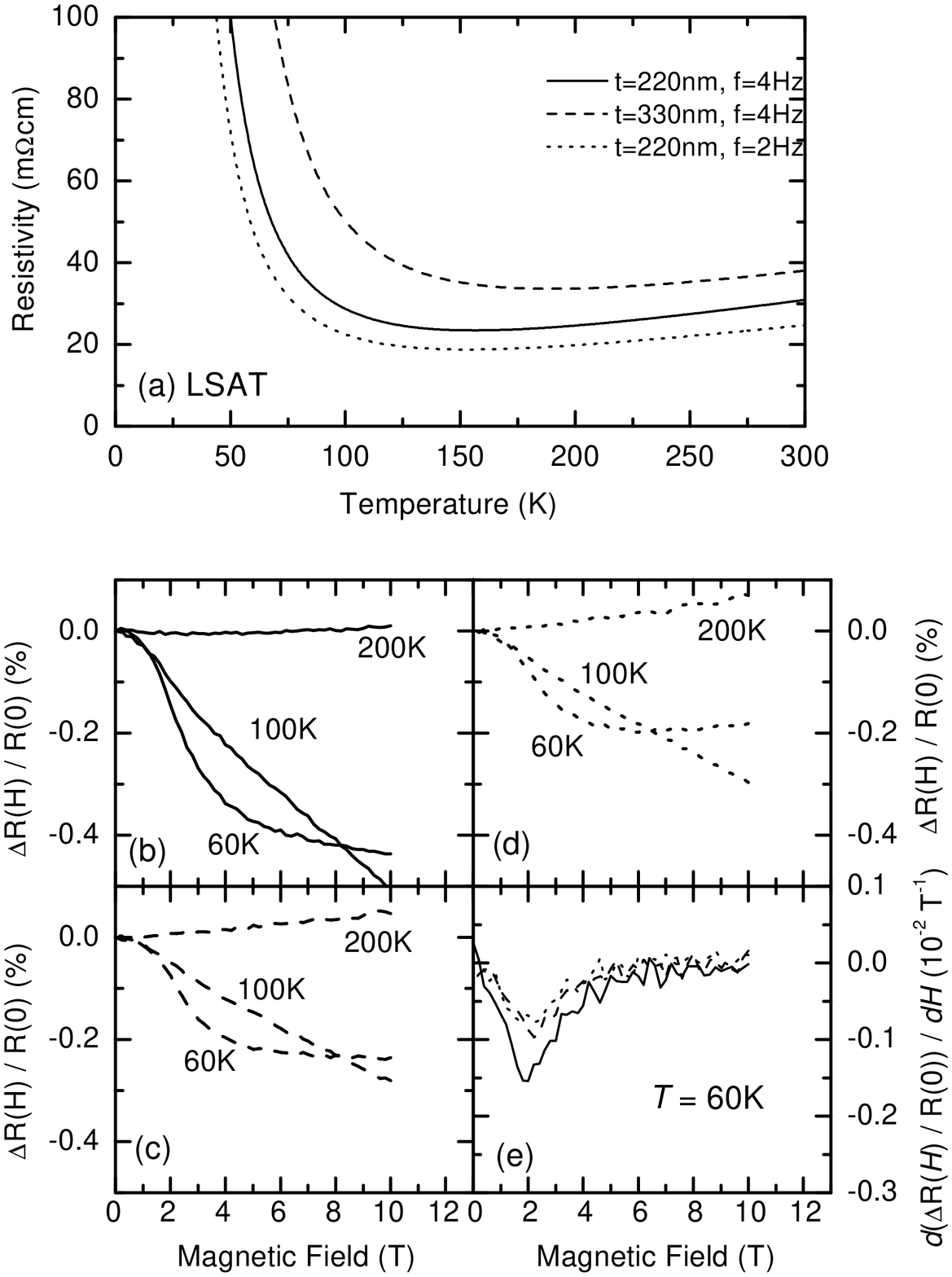}
\caption{
(a) Temperature dependence of the in-plane resisitivty of 
La${}_{1.99}$Sr${}_{0.01}$CuO${}_4$ thin films on LSAT (100) 
prepared under the conditions different from those shown in 
Figs.~\protect\ref{Fig.2} and \protect\ref{Fig.3}. 
(b) The magnetoresistance of the film ($t$ = 2200{\AA}) prepared 
at $f$ = 4Hz. 
(c) The magnetoresistance of the film ($t$ = 3300{\AA}) prepared 
at $f$ = 4Hz. 
(d) The magnetoresistance of the film ($t$ = 2200{\AA}) prepared 
at $f$ = 2Hz. 
(e) The first derivative of the magnetoresistance at $T$= 60K 
shown in (b)-(d). 
}
\label{Fig.6}
\end{figure}

One may wonder if the result obtained above is strongly dependent 
on film thickness, because the lattice strain applied from substrate may 
be relaxed by increasing the film thickness in general, 
which will result a recovery of step-like magnetoresistance. 
Since the film grown on LSAT indicates slight but finite corrugations 
in the CuO${}_2$ plane as is implied from Fig.~\ref{Fig.3}, 
LSAT is considered to be the most appropriate substrate to study 
the thickness dependence. 
Figure~\ref{Fig.6} shows the resistivity and magnetoresistance of the films 
grown on LSAT (100) prepared under the condition different 
from that for the film shown in Fig.~\ref{Fig.4}: 
$t$=2200{\AA} at $f$=4~Hz, $t$=3300{\AA} at $f$=4~Hz, 
and $t$=2200{\AA} at $f$=2~Hz. 
Although the magnitude of the resistivity is not completely identical, 
its temperature dependence is similar to one another and that for typical 
insulating LSCO with $x$=0.01 as shown in Fig.~\ref{Fig.6}(a). 
\cite{Ando2} 
Figures~\ref{Fig.6}(b) - \ref{Fig.6}(d) show the magnetoresistance 
measured at $T$ = 200, 100, and 60~K. 
It is easily seen that all the films exhibit almost the 
same behavior as that in Fig.~\ref{Fig.4}(e). 
At $T$ = 200~K, the magnetoresistance is slightly positive and 
almost featureless. 
When temperature is decreased down to 100~K, the negative 
magnetoresistance becomes apparent. but step-like behavior is still absent. 
After the temperature is decreased to 60~K, the step-like behavior shows up. 
What we should emphasize is that the overall magnetoresistance behavior 
is qualitatively the same among Figs.~\ref{Fig.4}(e), \ref{Fig.6}(b)-\ref{Fig.6}(d). 
The magnitude of magnetoresistance at $H$ = 10~T is less than 0.5~{\%} 
at $T$ = 60~K, 
again suggesting the corrugation of the CuO${}_2$ plane remains 
but its tilting is strongly suppressed. 
The first derivative of the magnetoresistance traces almost 
the identical line (Fig.~\ref{Fig.6}(e)), 
which suggests the weak-ferromagnetic transition occurs at the same field 
regardless of the film thickness. 
As a result, we may safely conclude that the general behavior of the 
magnetoresistance shown in Fig.~\ref{Fig.4} is not so strongly dependent 
on the film thickness and/or the growth condition.

We are now ready to discuss the microscopic structure of the CuO${}_2$ plane. 
As has been reported for La${}_2$CuO${}_{4+\delta}$ 
(Ref.~4) and 
La${}_{1.99}$Sr${}_{0.01}$CuO${}_4$ 
(Ref.~5) 
single crystals, 
the magnetic field applied parallel to the $c$ axis induces a 
WF transition at the critical field $H_{WF}$ $\approx$ 4-5~T, which is 
accompanied by a large steplike negative magnetoresistance. 
Whatever the origin of this peculiar magnetoresistance is,
\cite{Shekhtman1,Ando2} 
the WF transition is the most clear evidence of a finite spin canting 
out of the CuO${}_2$ plane that is inherent in the corrugated structure 
of CuO${}_2$ planes characteristic of the LTO phase. 
The present results indicate that the film grown on YAP actually has 
a corrugation in the CuO${}_2$ plane similar to that in bulk crystals. 
The experimentally determined $H_{WF}$ (= 4~T) implies 
that the film grown on YAP is almost identical to bulk crystals, 
because single crystals show almost the same transition field. 
\cite{Ando2} 
However, it should be noted that the transition width across $H_{WF}$ 
is broader in this film than in reported crystals.
\cite{Thio1,Ando2} 
This suggests a spatial inhomogeneity of the transition field due to a finite 
lattice misfit between the crystal and the substrate.

The magnetoresistance of the films grown on LSAT and STO is interesting. 
As shown in Fig.~\ref{Fig.1}(e), the $c$-axis lengths of these films 
are shorter than that of the bulk crystals. 
This implies that the tensile strain is actually applied to the films, 
and thus we may expect the symmetry of CuO${}_2$ planes to become 
tetragonal following the substrate symmetry. 
On the other hand, as can be seen in Figs.~\ref{Fig.4}(a) and \ref{Fig.4}(b), 
a weakened but finite steplike magnetoresistance remains, 
indicating that the corrugation in the CuO${}_2$ plane survives. 
The most probable explanation is that these films are separted in two parts: 
lower fully-strained layer and upper relaxed layer, 
and the corrugation structure remaing in the relaxed layer is resposible for 
the steplike magnetoresistance. 
If lattice relaxation proceeds, the crystal will approach 
its original form with finite corrugation in the CuO${}_2$ plane. 
However, there is no sign of bulk orthorhombicity in these films 
as is demonstrated by Figs.~\ref{Fig.2}(d) and \ref{Fig.2}(e). 
In order to explain these results self consistently, 
we expect the film to have a microscopic twin structure: 
the films are separated into domains that are still orthorhombic 
but their orthorhombicity is not as large as in the films grown on YAP. 
In this case, the corrugation in the CuO${}_2$ plane is reasonably 
suppressed leading to the reduction in the WF moment. 
Certainly, the magnitude of the WF moment cannot be evaluated directly 
from the magnetoresistance. 
However, if the reduction in the WF moment really takes place, 
a substantial suppression of the effective interplane coupling 
can be deduced from the experimentally observed reduction of $H_{WF}$, 
because the product of $H_{WF}$ and the magnitude of the WF moment is 
roughly proportional to the effective inter-plane coupling. 
Such suppression of the interplane coupling is consistent with the scenario 
that the symmetry of CuO${}_2$ planes becomes less orthorhombic.

Our results also show an advantage of LSAO over other substrates. 
The weak ferromagnetism is more strongly suppressed in the film grown 
on LSAO than in films grown on LSAT or STO. 
This is probably because that the lattice misfit of LSCO to LSAO 
is smaller than those to STO and LSAT.
\cite{misfit}
We emphasize that not only the tensile strain but also the compressive 
strain is helpful in removing the corrugation from the CuO${}_2$ plane. 
This cannot be simply understood because the presence of corrugation 
implies that the CuO${}_2$ plane has already been compressed 
by a rather small La${}_2$O${}_2$ blocking layer. 
With respect to this point, our results strongly indicate that 
the substrate symmetry is also essential for the removal of corrugation, 
and that the tetragonal substrate works well for these particular LSCO 
thin films.

Finally, let us briefly discuss the AF long-range order 
in these films, because not only $H_{WF}$ but also $T_N$ seems to be 
affected by the epitaxial strain. 
Figures~\ref{Fig.4}(c)-\ref{Fig.4}(f) show the magnetoresistance 
measured at $T$ = 100~K. 
In the film grown on YAP and STO, we can find weak ferromagnetism 
at this temperature, 
indicating that $T_N$ of these films is higher than 100~K. 
On the other hand, the magnetoresistance data for the films grown 
on LASO and LSAT show no such features, 
suggesting that the system is still in the paramagnetic phase, 
and the latter two films probably have lower $T_N$'s than the former ones 
or the bulk crystals. 
We consider that the reduction in $T_N$ is also caused by the tetragonal 
symmetry that the substrate introduces to the film. 
If CuO${}_2$ planes are truly tetragonal and flat, 
the inter-plane exchange interactions between the nearest-neighbor Cu ions 
on the adjacent layers are strongly frustrated. 
Consequently, a strong reduction in the effective interplane interactions 
is expected, which should suppress the formation of the 
magnetic long-range order. 
Thus, the reduction in $T_N$ together with $H_{WF}$ also confirms that 
the films become less orthorhombic.

To summarize, we have found that the in-plane magnetoresistance 
of antiferromagnetic LSCO thin films is strongly dependent 
on the substrate material. 
The difference in the magnetoresistance behavior is attributed to 
the change in the corrugation structure of the CuO${}_2$ plane. 
Within our experiments, the flattest CuO${}_2$ plane is obtained 
in films grown on LaSrAlO${}_4$(001) substrates. 
It is suggested that not only the lattice parameters of the substrate 
but also its symmetry plays a significant role in removing 
the corrugations. 
We expect that a similar epitaxial strain works in superconducting 
LSCO thin films accounting for the observed $T_c$ enhancement.

The author thanks A. N. Lavrov, Seiki Komiya, and Yoichi Ando 
for stimulating discussions and also for a critical reading of the 
manuscript.



\begin{references}
\bibitem{Sato1} H. Sato and M. Naito, 
Physica C {\bf 274}, 221 (1997); 
H. Sato, A. Tsukada, M. Naito, and A. Matsuda, 
Phys. Rev. B {\bf 61}, 12447 (2000). 
\bibitem{Sato2} H. Sato, A. Tsukada, M. Naito, and A. Matsuda, 
Phys. Rev. B {\bf 62}, R799 (2000). 
\bibitem{Locquet1} J. -P. Locquet, J. Perret, J. Fompeyrine, 
E. M{\"a}chler, J. W. Seo, and G. Van Tendeloo, 
Nature (London) {\bf 394}, 453 (1998). 
\bibitem{Thio1} T. Thio, T. R. Thurston, N. W. Preyer, P. J. Picone, 
M. A. Kastner, H. P. Jenssen, D. R. Gabbe, C. Y. Chen, R. J. Birgeneau, 
and A. Aharony, 
Phys. Rev. B {\bf 38}, 905 (1988). 
\bibitem{Ando2} Y. Ando, A. N. Lavrov, and S. Komiya, 
Phys. Rev. Lett. {\bf 90}, 247003 (2003). 
\bibitem{Lavrov2} A. N. Lavrov, I. Tsukada, and Y. Ando, 
Phys. Rev. B {\bf 68}, 094506 (2003). 
\bibitem{Lavrov1} A. N. Lavrov, Y. Ando, S. Komiya, and I. Tsukada, 
Phys. Rev. Lett. {\bf 87}, 017007 (2001). 
\bibitem{Bozovic1} I. Bozovic, G. Logvenov, I. Belca, B. Narimbetov, 
and I. Sveklo, 
Phys. Rev.Lett. {\bf 89}, 107001 (2002). 
\bibitem{Radaelli1} P. G. Radaelli, D. G. Hinks, A. W. Mitchell, B. A. Hunter, 
J. L. Wagner, B. Dabrowski, K. G. Vandervoort, H. K. Viswanathan, 
and J. D. Jorgensen, 
Phys. Rev. B {\bf 49}, 4163 (1994). 
\bibitem{Suzuki1} T. Suzuki and T. Fujita, 
Physica C {\bf 159}, 111 (1989). 
\bibitem{Eckstein1} J. N. Eckstein, I. Bozovic, K. E. von Dessonneck, 
D. G. Schlom, J. S. Harris, Jr., and S. M. Baumann, 
Appl. Phys. Lett. {\bf 57}, 931 (1990). 
\bibitem{Ando1} Y. Ando, K. Segawa, S. Komiya, and A. N. Lavrov, 
Phys. Rev. Lett. {\bf 88}, 137005 (2002). 
\bibitem{Shekhtman1} L. Shekhtman, I. Ya. Korenblit, and A. Aharony, 
Phys. Rev. B {\bf 49}, 7080 (1994). 
\bibitem{misfit} We should calculate the misfit both along the $a$ and 
$b$ axis 
for LTO LSCO. By using the lattice parameters of bulk samples (Ref.[10]), 
the misfit is 0.85{\%} ($\parallel$ $a$) and 1.60{\%} 
($\parallel$ $b$) to LSAO, 
-2.10{\%} ($\parallel$ $a$) and -1.37{\%} ($\parallel$ $b$) to LSAT, 
and -2.99{\%} ($\parallel$ $a$) and -2.26{\%} ($\parallel$ $b$) to STO. 
The smallest misfit is achieved to LSAO both along $a$ and along $b$. 
\end{references}
\end{document}